\documentclass[3p,times,procedia]{elsarticle}
\usepackage{nupha_ecrc}
\usepackage{nicefrac}

\volume{00}

\firstpage{1}

\journalname{Nuclear Physics A}

\runauth{}

\jid{nupha}

\jnltitlelogo{Nuclear Physics A}

\usepackage{amssymb}
\usepackage[figuresright]{rotating}

\newcommand{\bea}{\begin{eqnarray}}
\newcommand{\eea}{\end{eqnarray}}
\newcommand{\bel}[1]{\begin{eqnarray}\label{#1}}
\newcommand{\eel}{\end{eqnarray}}

\newcommand{\p}{\partial}

\newcommand{\f}[2]{\frac{#1}{#2}}
\newcommand{\onehalf}{{\nicefrac{1}{2}}}

\newcommand{\tr}{{\rm tr}}

\def\g5{\gamma_5}

\def\TmnU{T^{\mu\nu}}
\def\TnmU{T^{\nu\mu}}

\def\oTmnU{{\widehat T}^{\mu\nu}}

\def\SmnU{{\Sigma}^{\mu\nu}}

\def\S0iU{{\Sigma}^{0i}}

\def\SlmnU{S^{\lambda, \mu\nu}}
\def\SmlnU{S^{\mu, \lambda\nu}}

\def\oSmlnU{{{\widehat S}^{\mu, \lambda\nu}}}

\def\wj{{\widehat j}}

\def\n0{n_{(0)}}
\def\e0{\varepsilon_{(0)}}
\def\P0{P_{(0)}}

\def\TmnU{T^{\mu\nu}}                      % energy-momentum tensor

\def\rhoLEQ{{\widehat{\rho}}_{\rm \small LEQ}}

\def\Wpmxk{{\cal W}^{\pm}(x,k)}

\def\Weqpxk{{\cal W}^{+}_{\rm eq}(x,k)}

\def\Fpmxk{{\cal F}^{\pm}(x,k)}

\def\Ppmxk{{\cal P}^{\pm}(x,k)}

\begin{document}

\begin{frontmatter}



%% Instructions from Editor: Please use the following \dochead only in the preprint version (e-print arXiv etc.); 
%% use empty \dochead{} when submitting to Nuclear Physics A!
\dochead{XXVIIIth International Conference on Ultrarelativistic Nucleus-Nucleus Collisions\\ (Quark Matter 2019)}
%% e.g. \dochead{17th International Conference on Dynamical Processes in Excited States of Solids}

\title{Hydrodynamics of massive particles with spin $\onehalf$}


\author{Wojciech Florkowski}

\address{Institute of Theoretical Physics, Jagiellonian University \\
ul. prof. St. {\L }ojasiewicza 11, 30-348 Kraków, Poland}

\begin{abstract}
The formulation of relativistic hydrodynamics for massive particles with spin $\onehalf$ is shortly reviewed. The proposed framework is based on the Wigner function treated in a semi-classical approximation or, alternatively, on a classical treatment of spin $\onehalf$. Several theoretical issues regarding the choice of the energy-momentum and spin tensors used to construct the hydrodynamic framework with spin are discussed in parallel.
\end{abstract}


\end{frontmatter}

% \linenumbers

\section{Introduction}
\label{}

Non-central heavy-ion collisions make it possible that large amount of the initial orbital angular momentum is transferred to produced systems. Some part of such an angular momentum can be subsequently shifted from the orbital part to the spin part. The latter can be reflected in the spin polarization of  produced hadrons such as $\Lambda$ and  $\bar{\Lambda}$ hyperons~\cite{Voloshin:2004ha,Liang:2004ph,Betz:2007kg,Gao:2007bc,Voloshin:2017kqp}. The spin polarization of  $\Lambda$'s and  $\bar{\Lambda}$'s has been indeed measured by the STAR Collaboration at BNL~\cite{STAR:2017ckg, Adam:2018ivw} and the data shows global, out-of-plane polarization, which reminds us of the Einstein~--~de~Haas and Barnett effects~\cite{dehaas:1915,RevModPhys.7.129}. 

The phenomenon of global polarization has been successfully explained by the hydrodynamic models, which directly identify spin polarization effects with the so-called thermal vorticity~\cite{Karpenko:2016jyx}. The latter is defined  by the rank-two antisymmetric tensor $\varpi_{\mu \nu} = -\frac{1}{2} (\partial_\mu \beta_\nu-\partial_\nu \beta_\mu)$, where $\beta_\mu$ is the ratio of the hydrodynamic flow $u_\mu$ and local temperature $T$, $\beta_\mu = u_\mu/T$~\cite{Becattini:2007nd,Becattini:2009wh,Becattini:2016gvu}. There exist, however, problems regarding description of the longitudinal polarization, since a theoretically predicted longitudinal polarization of $\Lambda$'s~\cite{Becattini:2017gcx} has an opposite sign of the dependence on the azimuthal angle of emitted particles, as compared to the data~\cite{Niida:2018hfw}. 
 
Using general thermodynamic arguments  one can argue that there are situations where the spin polarization effects (quantified by the tensor $\omega_{\mu\nu}$, dubbed below the spin polarization tensor) are independent of the thermal vorticity  $\varpi_{\mu \nu}~$\cite{Becattini:2018duy}. Associated with this generalization is an extension of the concept of local thermodynamic equilibrium (where $\omega_{\mu\nu}=\varpi_{\mu \nu}$) to a local spin-thermodynamic equilibrium (where we allow for $\omega_{\mu\nu} \neq \varpi_{\mu \nu}$). This idea was originally proposed in Ref.~\cite{Florkowski:2017ruc} and developed in Refs.~\cite{Florkowski:2017dyn,Florkowski:2018myy,Florkowski:2018ahw} (see the recent review~\cite{Florkowski:2018fap} and related works~\cite{Sun:2018bjl,Weickgenannt:2019dks,Gao:2019znl}).  Within this approach, the space-time evolution of polarization is determined by the conservation law for total angular momentum --- for particles with spin, the latter turns out to have a non-trivial form. 
  
Incorporation of spin polarization into the hydrodynamic framework is an interesting challenge, as the present works say little about the changes of the spin polarization during the heavy-ion collision process. As the fluid dynamics represents now the basic element of heavy-ion models (for recent developments see \cite{Florkowski:2017olj,Romatschke:2017ejr}), it is even more demanding to have spin effects included into the hydrodynamic picture of heavy-ion collisions. So far, relatively little work has been done in this direction, although the studies of fluids with spin have a rather long history that started in 1940s \cite{Weyssenhoff:1947iua,Bohm:1958,Halbwachs:1960aa}. 

\section{Local spin-thermodynamic equilibrium}

The local spin-thermodynamic equilibrium is described, besides the standard hydrodynamic quantities such as temperature $T(x)$, flow four-vector $u^\mu(x)$, and chemical potential $\mu(x) = \xi(x) T(x)$, by the spin chemical potential $\Omega_{\mu\nu}(x) = \omega_{\mu\nu}(x) T(x)$~\cite{Becattini:2018duy}. One uses these quantities to construct the density operator $\rhoLEQ$ and to obtain the expectation values of the energy-momentum tensor $\TmnU$, the spin tensor $\SmlnU$, and the baryon current $j^{\mu}$:
\bea
\TmnU =  \tr \left(\rhoLEQ \, \oTmnU \right), \,\,
\SmlnU = \tr \left(\rhoLEQ \,  \oSmlnU \right), \,\,
j^{\mu} = \tr \left(\rhoLEQ \, \wj^{\mu}\right).
\label{eq:rh0}
\eea
In this way we obtain constitutive equations:
\bea
\TmnU = \TmnU [\beta,\omega,\xi], \,\,
\SmlnU = \SmlnU [\beta,\omega,\xi], \,\,
j^\mu = j^\mu [\beta,\omega,\xi].
\label{eq:TSj}
\eea
With dissipation effects neglected, one can assume that the density operator $\rhoLEQ$ is constant, which leads to the following equations:
\bea
\p_\mu \TmnU = 0, \quad
\p_\lambda \SlmnU = \TnmU -\TmnU, \quad
\p_\mu  j^\mu =0.
\label{eq:h1}
\eea
These are eleven equations for eleven unknown functions (temperature, three independent components of the fluid four-velocity, chemical potential, and six independent components of the tensor $\omega_{\mu\nu}$, which becomes now a new hydrodynamic quantity). We note that in general only the total angular momentum is conserved, which leads to the middle equation in (\ref{eq:h1}).

%%%%%%%%%%%%%%%%%%%%%%%%%%%%%%%%%%%%%%%%%
\section{Semiclassical kinetic equation}

Combining the concepts presented in \cite{Becattini:2013fla} with the idea of spin-thermodynamic equilibrium, one can introduce
equilibrium Wigner functions. Using a semi-classical formalism worked out in~\cite{DeGroot:1980dk}, we write
\bea
\Weqpxk = \frac{1}{2} \sum_{r,s=1}^2 \int \frac{d^3p}{(2\pi)^3 E_p}\,
\delta^{(4)}(k-p) u^r(p) {\bar u}^s(p) f^+_{rs}(x,p),
\nonumber
\eea
where  $k$ is the four-momentum and $f^+_{rs}(x,p)$ is the spin-dependent phase-space density matrix of particles~\cite{Leader:2001}, which in equilibrium depends on $\beta^\mu$, $\omega_{\mu\nu}$, and $\xi$. A similar expression can be introduced for antiparticles. Any Wigner function can be furthermore expressed as a linear combination of the generators of the Clifford algebra~\cite{Elze:1986hq,Elze:1986qd,Vasak:1987um,Zhuang:1995pd,Florkowski:1995ei,Gao:2017gfq},
\bea
&& \Wpmxk =  \f{1}{4} \left[ \Fpmxk + i \gamma_5 \Ppmxk + \gamma^\mu {\cal V}^\pm_{ \mu}(x,k) + \gamma_5 \gamma^\mu {\cal A}^\pm_{ \mu}(x,k)
+ \SmnU {\cal S}^\pm_{ \mu \nu}(x,k) \right],   \label{eq:W}
\eea
which is a very useful expression for studying a semiclassical limit of the quantum kinetic equation,
\bel{eq:eqforWC}
\left(\gamma_\mu K^\mu - m \right) {\cal W}(x,k) = C[{\cal W}(x,k)].
\label{eq:kineq}
\eel
Here  $K^\mu$ is the operator defined by the expression
\bel{eq:K}
K^\mu = k^\mu + \frac{i \hbar}{2} \,\p^\mu \label{eq:Kmu}
\eel
and $C[{\cal W}(x,k)] $ is the collision term. In global or local equilibrium, the collision term vanishes and
one can study only the left-hand side of (\ref{eq:kineq}) that should be equal to zero in this case.

At this point it is important to distinguish between the global and local equilibrium. In the case of global equilibrium the Wigner function ${\cal W}(x,k)$ exactly satisfies the equation
\bel{eq:eqforW}
\left(\gamma_\mu K^\mu - m \right) {\cal W}(x,k) = 0.
\label{eq:kineq0}
\eel
From the leading and next-to-leading terms of the expansion of $ {\cal W}$ in powers of $\hbar$, one finds that $\xi$~=~const., $\omega_{\mu\nu}$~=~const., and $\partial_\mu \beta_\nu - \partial_\nu \beta_\mu = 0$. The last equation is known as the Killing equation that (in the flat space-time) has a solution $\beta_\mu = b^0_\mu + \omega^0_{\mu \nu} x^\nu$ with  $b^0_\mu$~=~const. and $\omega^0_{\mu \nu} = - \omega^0_{\nu \mu} $~=~const. Interestingly, the tensors $\omega_{\mu\nu}$ and $\omega^0_{\mu\nu} = \varpi_{\mu\nu}$ might be different.

In the case of local equilibrium, one assumes that only specific moments of Eq.~(\ref{eq:kineq0}) vanish. This point has been discussed in more detail in Ref.~\cite{Florkowski:2018ahw}, where it is shown that this procedure leads to the following equations:
\bea
\p_\mu j^\mu_{\rm GLW}(x)  = 0, \quad \p_\alpha T^{\alpha\beta}_{\rm GLW}(x) = 0, \quad
\p_\lambda S^{\lambda , \mu \nu }_{\rm GLW}(x) = 0.
\label{eq:GLWhydro}
\eea
This form is consistent with the general scheme of the hydrodynamics with spin discussed above, however, the forms of the tensors appearing in Eqs.~(\ref{eq:GLWhydro}) are different from those used in a phenomenological approach of Ref.~\cite{Florkowski:2017ruc}. As a matter of fact,  these forms agree with the expressions used by de Groot, van Leeuwen, and van Weert in Ref.~\cite{DeGroot:1980dk}. This fact is stressed by the use of the GLW labels in Eqs.~(\ref{eq:GLWhydro}). We note that the first  numerical solutions of Eqs.~(\ref{eq:GLWhydro}) have been obtained recently in Ref.~\cite{Florkowski:2019qdp}. We also note that Eqs.~(\ref{eq:GLWhydro}) can be derived from an approach where the spin is treated in a classical way~\cite{Florkowski:2018fap}.

%%%%%%%%%%%%%%%%%%%%%%%%%% 
\section{Pseudo-gauge symmetry}

It is important to emphasize that one can also use the canonical versions of the energy-momentum and spin tensors to construct the hydrodynamic equations with spin. It turns out, that the canonical and GLW forms are connected by a pseudo-gauge transformation. Indeed, if we introduce a tensor $\Phi_{\rm can}^{\lambda, \mu\nu}$ defined by the relation
\bel{Phi}
\Phi_{\rm can}^{\lambda, \mu\nu} 
\equiv S^{\mu , \lambda \nu }_{\rm GLW}
-S^{\nu , \lambda \mu }_{\rm GLW},
 \label{eq:Phi}
\eel
one can check that
\bel{psg1GLW}
S^{\lambda , \mu \nu }_{\rm can}= S^{\lambda , \mu \nu }_{\rm GLW} -\Phi_{\rm can}^{\lambda, \mu\nu}, \qquad
T^{\mu\nu}_{\rm can} = T^{\mu\nu}_{\rm GLW} + \frac{1}{2} \p_\lambda \left(
\Phi_{\rm can}^{\lambda, \mu\nu}
+\Phi_{\rm can}^{\mu, \nu \lambda} 
+ \Phi_{\rm can}^{\nu, \mu \lambda} \right) .
\eel
The pseudo-gauge transformation given above is similar to the Belinfante construction but it does not eliminate the spin tensor which can be used to describe spin degrees of freedom. 

One can check that the canonical and GLW hydrodynamic equations are the same, however, the forms of the energy-momentum and spin tensors are different. In particular, in the canonical case the energy-momentum tensor is not symmetric and, consequently, the divergence of the canonical spin tensor does not vanish. The GLW version seems to be a convenient rearrangements of the terms used to define $T^{\mu\nu}$ and $S^{\lambda , \mu \nu }$, which leads to a symmetric $T^{\mu\nu}$ and a conserved $S^{\lambda , \mu \nu }$. We must admit, however, that it is not clear if such a rearrangement is possible if one goes beyond the semi-classical description discussed in this contribution.

%%%%%%%%%%%%%%%%%%%%%%%%%% 
\section{Conclusions}

The results presented in this contribution describe dynamics of a~perfect fluid consisting of massive particles with spin \nicefrac{1}{2}. 
The main challenge for next developments is the proper inclusion of dissipation (for example, a calculation of  kinetic coefficients related to spin observables). First steps in this direction have been made, for example, in Ref.~\cite{Hattori:2019lfp,Bhadury:2020puc}. In the closest future, it would be interesting to examine more closely the relation of the results presented in Ref.~\cite{Hattori:2019lfp} to the formalism discussed herein. It is also mandatory to study in more detail the relation between spin polarization and thermal vorticity. An effect describing convergence of the spin polarization tensor to the thermal vorticity should be included in the complete formalism of viscous hydrodynamics with spin.

\medskip
I thank F. Becattini, S. Bhadury, B. Friman, A. Jaiswal, A. Kumar, E. Speranza, R. Ryblewski for very fruitful collaboration and numerous illuminating discussions. This work was supported in part by the Polish National Science Center Grant  No. 2016/23/B/ST2/00717.

\bibliographystyle{elsarticle-num}
\bibliography{<your-bib-database>}



\end{document}